\documentclass[12pt]{article}
\usepackage{amssymb,amsmath,epsfig}
\usepackage{graphicx}
\allowdisplaybreaks
\begin{document}

\title{\bf Complexity Analysis of Dynamical Cylinder in Massive Brans-Dicke Gravity}
\author{M. Sharif \thanks{msharif.math@pu.edu.pk} and Amal Majid
\thanks{amalmajid89@gmail.com}\\
Department of Mathematics, University of the Punjab,\\
Quaid-e-Azam Campus, Lahore-54590, Pakistan.}

\date{}
\maketitle
\begin{abstract}
In this paper, a complexity factor is devised for a non-static
cylindrical system in the framework of massive Brans-Dicke theory.
The definition of complexity is developed by taking into account the
essential physical characteristics (such as anisotropy,
inhomogeneity, etc.) of the system. In order to determine the
complexity factor of the self-gravitating object, we acquire
structure scalars from the orthogonal splitting of the Riemann
tensor. Moreover, we discuss two patterns of evolution and choose
the homologous mode as the simplest pattern under the influence of
massive scalar field. We derive solutions in the absence as well as
presence of heat dissipation for a specific form of the scalar
field. The factors that induce complexity in an initially
complexity-free system are also examined. It is concluded that the
massive scalar field as well as heat dissipation contribute to the
complexity of the celestial system. Thus, a dynamical cylinder is
more complex as compared to its static counterpart.
\end{abstract}

\section{Introduction}

Self-gravitating cosmic structures provide significant information
regarding the origin and evolution of the universe. For this
purpose, astrophysicists have performed detailed surveys of these
large-scale structures (Large Synoptic Survey Telescope, Sloan
Digital Sky Survey, Two-degree Field Galaxy Redshift Survey) to
investigate the mechanism of the cosmos. However, the physical
features of their intricate interior regions depend on different
factors such as mass and matter composition. A slight fluctuation in
any of the matter variables may lead to a fundamental change in the
behavior of celestial systems. Therefore, it is necessary to define
a complexity factor that effectively represents the complicated
nature and physics of astrophysical components. The formulation of
this factor is based on the inter-relationship of state parameters
(mass, pressure, density, etc.) which helps in determining the
extent to which internal or external disturbances influence the
characteristics of the cosmic system. Such a factor also provides a
comparison of complexities in different astrophysical objects
through a stability criterion.

The concept of complexity has been explored in different avenues but
researchers have failed to agree upon a standard definition
\cite{1}. The notion of complexity first arose when structures of
two physical models (ideal gas and perfect crystal) were compared.
The molecules of ideal gas frequently change their positions while
the perfect crystal has a symmetric arrangement in which atoms
occupy fixed places. Thus, maximum information is required to
specify a probable state of ideal gas whereas a perfect crystal is
completely specified by the least amount of information. Despite the
differences in the atomic arrangements, the two models are allotted
zero complexity. The previous definitions accommodated the concepts
of information regarding atomic arrangement and symmetries as well
as quantification of geometrical attributes. Lopez-Ruiz et al.
\cite{5} also examined the complexity of self-gravitating structures
by examining their disequilibrium. They detected the differences
between probable states of the matter distribution and the
equiprobable configuration of the system. Researchers used energy
density instead of probability distribution in the definition to
examine the complexity of compact dense objects like white dwarfs
and neutron stars \cite{9}.

The definition proposed by Lopez-Ruiz et al. is considered
incomplete as it only encompasses the density of the matter source
and the contribution of significant state variables is neglected.
The astrophysical study of celestial systems has revealed that the
constituent particles are compactly arranged within the internal
configurations of dense stellar objects. Consequently, the radial
motion of particles is limited which generates anisotropy in
pressure \cite{2}. Other factors of anisotropic interiors include
pion condensation \cite{6}, phase transition \cite{5a} and
superfluid \cite{4}. Thus, anisotropy is an essential feature of
compact distributions. Herrera \cite{13} incorporated anisotropy as
well as energy density to determine the complexity of a static
sphere in the context of general relativity (GR). He put forward the
new definition on the assumption that an isotropic and homogeneous
distribution has zero complexity. According to this approach, an
effective measure of complexity was devised via splitting the
Riemann tensor in terms of structure scalars.

Herrera et al. \cite{13*} computed the structure scalars and
formulated the complexity factor for a homologously evolving sphere.
This notion of complexity was also extended to axially symmetric
cosmic objects and three complexity factors were evaluated through
the orthogonal splitting of the Riemann tensor \cite{15*}. In the
same work, a possible relation between complexity and symmetry of
the spacetime was explored. Herrera and his collaborators \cite{15a}
also established a hierarchy from the complex fluids to the simpler
Minkowski spacetime. Sharif and Butt \cite{14*} adopted Herrera's
approach to examine the impact of charge on the complexity of a
static sphere and deduced that the electromagnetic field enhances
the complexity of a system. Sharif and Tariq \cite{14c} investigated
the complex structure of a charged spherical system evolving in a
homologous pattern. Recently, the conditions which led to a
complexity-free system following a quasi-homologous mode of
evolution were also determined \cite{15b}.

The solution of vacuum cylindrical spacetime by Levi-Civita
motivated researchers to explore different astrophysical phenomena
with cylindrical geometry. The study of strong gravitational waves
emitted at the end of gravitational collapse of a stellar object
also attracted astrophysicists to examine salient features and
behavior of cylindrical cosmic bodies \cite{16*}. Herrera et al.
\cite{17a} inspected the regularity of static cylinders and deduced
that a spacetime matched to the Levi-Civita exterior does not admit
conformally flat solutions. The phenomenon of gravitational collapse
was discussed for a non-static cylinder through junction conditions
and it was found that radial pressure vanishes at the boundary of
the object \cite{17b}. Sharif and Abbas \cite{17c} considered a
charged radiating cylinder and computed the gravitational mass of
the anisotropic setup. Herrera et al. \cite{17d} specified the
fundamental properties and structure of an evolving self-gravitating
cylinder through structure scalars. Recently, complexity factors
corresponding to uncharged \cite{14a} as well as charged \cite{14b}
cylindrical systems were also developed.

The modified theories of relativity, obtained by modifying the
Einstein-Hilbert action, have opened the pathway to new avenues of
cosmological and astronomical phenomena. Moreover, these
modifications are aimed at providing solutions to two cosmic
problems: cosmic coincidence and fine-tuning. Scalar-tensor theories
include a wide range of modified theories obtained via modification
in the geometric structure of GR. Brans-Dicke (BD) theory, the
prototype of this family of modifications, is based on Machian
principle and Dirac hypothesis \cite{15}. In BD gravity, a dynamical
scalar field ($\psi(t)=\frac{1}{G(t)}$) acts as a mediator of
gravity while a coupling parameter $(\omega_{BD})$ gauges the
influence of the massless scalar field on matter distribution.
Larger values of scalar field correspond to smaller values of
coupling parameter.

The rapid expansion of the universe in the inflationary era is
effectively described by a large scalar field. Thus, small values of
coupling parameter efficiently explain the inflation of the cosmos
\cite{17} while the weak-field tests are consistent with BD theory
for larger values of $\omega_{BD}$ \cite{16}. This discrepancy is
resolved through a self-interacting potential $(V(\Phi))$ which
establishes a standard domain for the parameter through a
restriction on the mass of the scalar field ($\Phi$). Brans-Dicke
theory incorporating a potential function is termed as massive BD
(MBD) gravity. Sharif and Manzoor analyzed the effect of the massive
scalar field on spherical \cite{22} as well as cylindrical
\cite{23'} configurations by formulating structure scalars in the
background of MBD theory. Recently, Herrera's definition was adopted
to check the variation in the complexity of different configurations
under the influence of massive scalar field \cite{100, 100a}.
Researchers have utilized Herrera's definition in other modified
theories to construct complexity factors for different geometries as
well \cite{23*}.

This paper focuses on devising a complexity factor for a non-static
radiating cylindrical self-gravitating structure in the context of
MBD theory. The paper is arranged in the following format. In the
next section, the MBD field equations and relations between
different physical aspects of the dissipative cylinder are
established. We split the Riemann tensor to derive structure scalars
in section \textbf{3}. The pattern of evolution is discussed in
section \textbf{4}. In section \textbf{5}, solutions for
non-dissipative and dissipative scenarios are obtained by evaluating
kinematical as well as dynamical quantities. We inspect the
stability of the zero complexity condition in section \textbf{6}.
Section \textbf{7} provides a summary of the main results.

\section{Massive Brans-Dicke Theory and Matter Variables}

The MBD field equations (in relativistic units) are obtained by
varying the action
\begin{equation}\label{0}
S=\int\sqrt{-g}(\mathcal{R}\Phi-\frac{\omega_{BD}}{\Phi}\nabla^{\lambda}\nabla_{\lambda}\Phi
-V(\Phi)+\emph{L}_m)d^{4}x,
\end{equation}
with respect to the metric tensor as
\begin{eqnarray}\label{1}
G_{\lambda\mu}&=&T^{\text{(\text{eff})}}_{\lambda\mu}=\frac{1}{\Phi}(T_{\lambda\mu}^{(m)}
+T_{\lambda\mu}^\Phi).
\end{eqnarray}
Here the matter Lagrangian and Ricci scalar are represented by
$\emph{L}_m$ and $\mathcal{R}$, respectively while
$g=|g_{\lambda\mu}|$. Moreover, the energy-momentum tensor
$T_{\lambda\mu}^{(m)}$ specifies the matter source whereas
$T_{\lambda\mu}^\Phi$ incorporates the effects of the massive scalar
field as
\begin{equation}\label{3}
T_{\lambda\mu}^\Phi=\Phi_{,\lambda;\mu}-g_{\lambda\mu}\Box\Phi+\frac{\omega_{BD}}{\Phi}
(\Phi_{,\lambda}\Phi_{,\mu}
-\frac{g_{\lambda\mu}\Phi_{,\alpha}\Phi^{,\alpha}}{2})-\frac{V(\Phi)g_{\lambda\mu}}{2},
\end{equation}
where $\Box\Phi=\Phi^{,\lambda}_{~;\lambda}$. The equation of motion
for the scalar field is derived via action (\ref{0}) as
\begin{eqnarray}\label{2}
\Box\Phi=\frac{T^{(m)}}{3+2\omega_{BD}}+\frac{1}{3+2\omega_{BD}}
(\Phi\frac{dV(\Phi)}{d\Phi}-2V(\Phi)),
\end{eqnarray}
where $T^{(m)}=g_{\lambda\mu}T^{(m)\lambda\mu}$. We consider a
cylindrical cosmic object bounded by a hypersurface $\Sigma$ and
defined by the following line element
\begin{equation}\label{4}
ds^2=-X^2(t,r)dt^2+Y^2(t,r)dr^2+Z^2(t,r)(d\theta^2+dz^2).
\end{equation}
We assume that the cylinder is filled with anisotropic fluid
dissipating in the form of heat flux ($\mathfrak{q}$).  The radial
($p_r$)/transverse ($p_\perp$) pressures and energy density ($\rho$)
of the matter distribution are determined by the following
energy-momentum tensor
\begin{equation}\nonumber
T_{\lambda\mu}^{(m)}=(\rho+p_\perp)\mathfrak{u}_{\lambda}\mathfrak{u}_{\mu}+p_\perp
g_{\lambda\mu}+(p_r-p_\perp)\mathfrak{s}_\lambda
\mathfrak{s}_\mu+\mathfrak{q}_\lambda
\mathfrak{u}_\mu+\mathfrak{u}_\lambda \mathfrak{q}_\mu,
\end{equation}
where the heat flux ($\mathfrak{q}_\lambda=(0,\mathfrak{q}Y,0,0)$),
radial 4-vector ($\mathfrak{s}_\lambda=(0,Y,0,0)$) and 4-velocity
($\mathfrak{u}_\lambda=(-X,0,0,0)$) satisfy the following relations
\begin{eqnarray*}
\mathfrak{s}^\lambda \mathfrak{u}_\lambda=0,\quad
\mathfrak{s}^\lambda \mathfrak{s}_\lambda=1,\quad
\mathfrak{u}^\lambda \mathfrak{u}_\lambda=-1,\quad
\mathfrak{u}^\lambda \mathfrak{q}_\lambda=0.
\end{eqnarray*}

The energy-momentum tensor can be rewritten in a simplified form in
terms of the quantities
$\Pi_{\lambda\mu}=\Pi(\mathfrak{s}_{\lambda}\mathfrak{s}_{\mu}
-\frac{h_{\lambda\mu}}{3}),~ P=\frac{1}{3}(p_{r}+2p_{\perp}),~
\Pi=p_r-p_\perp,~
h_{\lambda\mu}=g_{\lambda\mu}+\mathfrak{u}_{\lambda}\mathfrak{u}_{\mu}$
as
\begin{equation}\label{5a}
T_{\lambda\mu}^{(m)}=\rho
\mathfrak{u}_{\lambda}\mathfrak{u}_{\mu}+Ph_{\lambda\mu}+\Pi_{\lambda\mu}+\mathfrak{q}(\mathfrak{s}_\lambda
\mathfrak{u}_\mu+\mathfrak{u}_\lambda \mathfrak{s}_\mu).
\end{equation}
The field equations are obtained through Eqs.(\ref{1})-(\ref{5a}) as
\begin{eqnarray}\label{6}
\frac{1}{\Phi}(X^2\rho-T_{00}^\Phi)&=&\frac{\dot{Z}}{Z}
\left(\frac{2\dot{Y}}{Y}+\frac{\dot{Z}}{Z}\right)-\frac{X^2}{Y^2}
\left(\frac{Z'^2}{Z^2}-\frac{2Y'Z'}{YZ}
+\frac{2Z''}{Z}\right),\\\label{7}
\frac{1}{\Phi}(-qXY+T_{01}^\Phi)&=&\frac{2 X'\dot{Z}}{XZ}+\frac{2
\dot{Y}Z'}{YZ}-\frac{2\dot{Z}'}{Z},\\\label{8}
\frac{1}{\Phi}(Y^2p_r+T_{11}^\Phi)&=&-\frac{Y^2}{X^2}\left(\frac{2
\ddot{Z}}{Z}-\frac{\dot{Z}\left(\frac{2
\dot{X}}{X}-\frac{\dot{Z}}{Z}\right)}{Z}\right)+\frac{Z'}{Z}
\left(\frac{2X'}{X}+\frac{Z'}{Z}\right),\\\nonumber
\frac{1}{\Phi}(Z^2p_r+T_{22}^\Phi)&=&-\frac{Z^2}{X^2}
\left(-\frac{\dot{X}\left(\frac{\dot{Y}}{Y}
+\frac{\dot{Z}}{Z}\right)}{X}+\frac{\dot{Y}\dot{Z}}{YZ}
+\frac{\ddot{Y}}{Y}+\frac{\ddot{Z}}{Z}\right)\\\label{9a}
&+&\frac{Z^2}{Y^2} \left(\frac{Z'
\left(\frac{X'}{X}-\frac{Y'}{Y}\right)}{Z}-\frac{X'Y'}{X
Y}+\frac{X''}{X}+\frac{Z''}{Z}\right),
\end{eqnarray}
where
\begin{eqnarray*}
T_{00}^\Phi&=&-\dot{\Phi}\left(\frac{\dot{Y}}{Y}+\frac{2
\dot{Z}}{Z}\right)+\frac{X^2\Phi'\left(-\frac{Y'}{Y}+\frac{2
Z'}{Z}\right)}{Y^2}+\frac{\omega_{BD}\left(\frac{X^2\Phi'^2}{Y^2}
+\dot\Phi^2\right)}{2\Phi}\\
&+&\frac{X^2\Phi''}{Y^2}+\frac{X^2}{2}V(\Phi),\\
T_{01}^\Phi&=&-\frac{X'\dot{\Phi}}{X}-\frac{\dot{Y}\Phi'}{Y}+\frac{\omega_{BD}}
{\Phi}\dot{\Phi}{\Phi}'+\dot{\Phi}',\\
T_{11}^\Phi&=&-\Phi'\left(\frac{X'}{X}+\frac{2
Z'}{Z}\right)-\frac{Y^2\dot{\Phi}\left(\frac{\dot{X}}{X}-\frac{2
\dot{Z}}{Z}\right)}{X^2}+\frac{\omega_{BD}\left(\frac{Y^2\dot{\Phi}^2}{X^2}
+\Phi'^2\right)}{2\Phi}\\
&+&\frac{Y^2\ddot{\Phi}}{X^2}-\frac{Y^2}{2} V(\Phi),\\\nonumber
T_{22}^\Phi&=&-\frac{Z^2\Phi'\left(\frac{X'}{X}-\frac{Y'}{Y}+\frac{Z'}{Z}\right)}
{Y^2}-\frac{Z^2\dot{\Phi}\left(\frac{\dot{X}}{X}-\frac{\dot{Y}}{Y}
-\frac{\dot{Z}}{Z}\right)}{X^2}-\frac{\omega_{BD}Z^2
\left(\frac{\Phi'^2}{Y^2}-\frac{\dot{\Phi}^2}{X^2}\right)}{2\Phi}\\
&+&\frac{Z^2
\ddot{\Phi}}{X^2}-\frac{Z^2\Phi''}{Y^2}-\frac{Z^2}{2}V(\Phi).
\end{eqnarray*}
Here, prime and dot represent differentiation with respect to $r$
and $t$, respectively. The conservation of dissipative fluid is
expressed in the form of following equations
\begin{eqnarray}\nonumber
&&\dot{T}_0^{0(\text{eff})}+(T_0^{0(\text{eff})}-T_1^{1(\text{eff})})\frac{\dot{Y}}{Y}
+2(T_0^{0(\text{eff})}-T_2^{2(\text{eff})})\frac{\dot{Z}}{Z}+(T_0^{1(\text{eff})})'
\\\label{100}
&&+(T_0^{1(\text{eff})})(\frac{X'}{X}+\frac{Y'}{Y}+2\frac{Z'}{Z})=0,\\\nonumber
&&\dot{T}_0^{1(\text{eff})}+(T_1^{1(\text{eff})})'+T_0^{1(\text{eff})}(\frac{\dot{X}}{X}
+\frac{\dot{Y}}{Y}+2\frac{\dot{Z}}{Z})-(T_0^{0(\text{eff})}-T_1^{1(\text{eff})})\frac{X'}{X}
\\\label{101}
&&+2(T_1^{1(\text{eff})}-T_2^{2(\text{eff})})\frac{Z'}{Z}=0.
\end{eqnarray}
Furthermore, the wave equation corresponding to the cylindrical
setup is given as
\begin{eqnarray}\nonumber
\Box\Phi&=&\frac{\Phi'\left(\frac{X'}{X}-\frac{Y'}{Y}+\frac{2
Z'}{Z)}\right)}{Y^2}-\frac{\dot{\Phi}\left(\frac{-\dot{X}}{X}
+\frac{\dot{Y}}{Y}+\frac{2\dot{Z}}{Z}\right)}{X^2}
-\frac{\ddot{\Phi}}{X^2}+\frac{\Phi''}{Y^2}\\\label{2*}
&=&\frac{1}{3+2\omega_{BD}}\left[-\rho+3P+
\left(\Phi\frac{dV(\Phi)}{d\Phi}-2V(\Phi)\right)\right].
\end{eqnarray}
In case of non-rotating fluid, the three kinematical variables
required to describe the motion of celestial system are
4-acceleration $(a_\lambda)$, expansion scalar $(\Theta)$ and shear
tensor $(\sigma_{\lambda\mu})$ which are, respectively defined as
\begin{equation}\nonumber
a_\lambda=\mathfrak{u}_{\lambda;\mu}\mathfrak{u}^{\mu}, \quad
\Theta=\mathfrak{u}^\lambda_{~;\lambda},\quad
\sigma_{\lambda\mu}=\mathfrak{u}_{\lambda;\mu}+a_{(\lambda}\mathfrak{u}_{\mu)}-\frac{1}{3}\Theta
h_{\lambda\mu}.
\end{equation}
The above quantities are evaluated corresponding to the cylindrical
setup as
\begin{eqnarray}\label{52}
a_1&=&\frac{X'}{X},\quad a^2=a_\lambda
a^\lambda=(\frac{X'}{XY})^2,\\\label{53}
\Theta&=&\frac{1}{X}(\frac{\dot{Y}}{Y}+2\frac{\dot{Z}}{Z}),\\\label{55}
\sigma_{11}&=&\frac{2}{3}Y^2\sigma,\quad
\sigma_{22}=-\frac{1}{3}Z^2\sigma,
\end{eqnarray}
with $a_\lambda=as_\lambda$ and
$\sigma=\sqrt{\frac{3}{2}\sigma^{\lambda\mu}\sigma_{\lambda\mu}}=\frac{1}{X}
(\frac{\dot{Y}}{Y}-\frac{\dot{Z}}{Z})$.

Thorne \cite{114} proposed the C-energy formula to calculate the
mass of a cylinder as
\begin{equation*}
m(t,r)=l\hat{E}=\frac{l}{8}\left(1-\frac{1}{l^2}
\nabla_{\lambda}\hat{r}\nabla^{\lambda}\hat{r}\right),
\end{equation*}
where
\begin{equation*}
\hat{r}=\mathfrak{R}l,\quad l^2=\chi_{(2)a}\chi_{(2)}^{a},\quad
\mathfrak{R}^2=\chi_{(1)a}\chi_{(1)}^{a}.
\end{equation*}
Here,
$\chi_1=\frac{\partial}{\partial\theta},~\chi_2=\frac{\partial}{\partial
z}$ and $\hat{E}$ is the gravitational energy per unit specific
length ($l$). The C-energy corresponding to Eq.(\ref{4}) is computed
as
\begin{equation}\label{9}
m(t,r)=\frac{Z}{2}(\frac{1}{4}+\frac{\dot{Z}^2}{X^2}-\frac{Z'^2}{Y^2}).
\end{equation}
The variation in energy of the anisotropic celestial object with
respect to time is evaluated via the proper time derivative
($D_T=\frac{1}{X}\frac{\partial}{\partial t}$) as
\begin{equation}\label{59}
D_Tm=-\frac{Z^2}{2}\left(T_{1}^{1(\text{eff})}+\frac{1}{4Z^2}\right)U
-\frac{T_{1}^{0(\text{eff})}}{2Y}Z^2E,
\end{equation}
where $E\equiv\frac{Z'}{Y}$. Moreover, the velocity of the
collapsing cylinder ($U=D_TZ<0$) is related to C-energy as
\begin{equation}\label{57}
E=\left(\frac{1}{4}+U^2-\frac{2m}{Z}\right)^\frac{1}{2}.
\end{equation}
It is noted that the term
$\left(T_{1}^{1(\text{eff})}+\frac{1}{4Z^2}\right)U$ in
Eq.(\ref{59}) contributes to the energy of the system if
$\left(T_{1}^{1(\text{eff})}+\frac{1}{4Z^2}\right)>0$. In order to
inspect the behavior of C-energy within the adjacent walls of the
cylinder we compute its proper radial derivative
$D_R=\frac{1}{Z'}\frac{\partial}{\partial r}$ (formulated from
circumference radius $R=Z$ of the cylinder within the hypersurface
\cite{101}) as
\begin{equation}\label{60}
D_Rm=\frac{Z^2}{2}\left(T_{0}^{0(\text{eff})}
+\frac{T_{1}^{0(\text{eff})}}{Y}\frac{U}{E}\right)-\frac{1}{4Z^2},
\end{equation}
which implies that if $\frac{T_{1}^{0(\text{eff})}}{Y}>0$ the energy
of the system decreases due to the second term within the brackets
as $U<0$. Integrating of Eq.(\ref{60}) provides
\begin{equation}\label{61}
\frac{3m}{Z^3}=-\frac{T_0^{0(\text{eff})}}{2}+\frac{1}{2Z^3}\int_0^r
Z'Z^3(D_RT_0^{0(\text{eff})}-
\frac{3T_{01}^{(\text{eff})}}{XYZ}\frac{U}{E})dr+\frac{3}{4Z^4}.
\end{equation}

The tidal forces experienced by a celestial system due to nearby
gravitational field play a vital role in determining its significant
physical features. In order to incorporate the effects of tidal
forces in our work, we evaluate the Weyl tensor defined as
\begin{equation}\label{1'}
C^{\lambda}_{\alpha\beta\sigma}=\mathcal{R}^{\lambda}_{\alpha\beta\sigma}-\frac{\mathcal{R}^{\lambda}_{\beta}}
{2}g_{\alpha\sigma}+\frac{\mathcal{R}_{\alpha\beta}}{2}\delta^{\lambda}_{\sigma}
-\frac{\mathcal{R}_{\alpha\sigma}}{2}\delta^{\lambda}_{\beta}
+\frac{\mathcal{R}^{\lambda}_{\sigma}}{2}g_{\alpha\beta}+\frac{1}{6}(\delta^{\lambda}_{\beta}
g_{\alpha\sigma}+g_{\alpha\beta}\delta^{\lambda}_{\sigma}),
\end{equation}
where $\mathcal{R}^{\lambda}_{\alpha\beta\sigma}$ and
$\mathcal{R}_{\alpha\beta}$ represent the Riemann and Ricci tensors,
respectively. The trace-free magnetic ($H_{\lambda\mu}$) and
electric ($\xi_{\lambda\mu}$) parts of the Weyl tensor are obtained
through $\mathfrak{u}^\gamma$ as
\begin{eqnarray}\label{11a}
H_{\lambda\mu}&=&\frac{1}{2}\eta_{\lambda\nu\epsilon\beta}
C_{\mu\gamma}~^{\epsilon\beta}\mathfrak{u}^{\nu}\mathfrak{u}^{\gamma},\\\label{11b}
\xi_{\lambda\mu}&=&C_{\lambda\beta\mu\sigma}\mathfrak{u}^{\beta}\mathfrak{u}^{\sigma}.
\end{eqnarray}
In case of cylindrical symmetry, the magnetic and electric parts are
non-vanishing in general. However, the magnetic part corresponding
to the considered setup vanishes whereas the electric part turns out
to be
\begin{equation}\label{11}
\xi_{\lambda\mu}=C_{\lambda\gamma\mu\delta}\mathfrak{u}^{\gamma}\mathfrak{u}^{\delta}=
\varepsilon(\mathfrak{s}_{\lambda}\mathfrak{s}_{\mu}+\frac{h_{\lambda\mu}}{3}),
\end{equation}
where
\begin{eqnarray}\nonumber
\varepsilon&=& \frac{\left(\frac{Z'}{Z}-\frac{X'}{X}\right)
\left(\frac{Y'}{Y}+\frac{Z'}{Z}\right)+\frac{X''}{X}-\frac{Z''}{Z}}{2Y^2}
+\frac{\frac{\ddot{Z}}{Z}-\frac{\ddot{Y}}{Y}
-\left(\frac{\dot{X}}{X}+\frac{\dot{Z}}{Z}\right)
\left(\frac{\dot{Z}}{Z}-\frac{\dot{Y}}{Y}\right)}{2X^2}.\\\label{14}
\end{eqnarray}
Furthermore, the impact of the massive scalar field on pressure,
Weyl tensor and energy density is demonstrated in the following
relation
\begin{equation}\label{15}
\frac{\partial}{\partial
t}[\varepsilon-\frac{1}{2}(-T_0^{0(\text{eff})}-T_1^{1(\text{eff})}+T_2^{2(\text{eff})})]
=\frac{3\dot{Z}}{Z}[\frac{1}{2}(-T_0^{0(\text{eff})}+T_2^{2(\text{eff})})
-\varepsilon]-\frac{3Z'}{2Z}T_0^{1(\text{eff})}).
\end{equation}

\section{Structure Scalars}

In this section, we formulate a complexity factor for the dynamical
object in terms of structure scalars which are determined via the
orthogonal splitting of Riemann tensor. The procedure of splitting
Riemann tensor was first applied by Herrera \cite{23} by expressing
the Riemann tensor in terms of trace and trace-free parts as follows
\begin{equation}\label{26}
\mathcal{R}^{\alpha\delta}_{\beta\gamma}=C^{\alpha\delta}_{\beta\gamma}+
2T^{(\text{eff})[\alpha}_{[\beta}\delta^{\delta]}_{\gamma]}+
T^{(\text{eff})}
\left(\frac{1}{3}\delta^{\alpha}_{[\beta}\delta^{\delta}_{\gamma]}-
\delta^{[\alpha}_{[\beta}\delta^{\delta]}_{\gamma]}\right).
\end{equation}
Employing the expressions
\begin{eqnarray}\nonumber
\mathcal{R}^{\alpha\delta}_{(I)\beta\gamma}&=&\frac{2}{\Phi}\left[\rho
\mathfrak{u}^{[\alpha}\mathfrak{u}_{[\beta}\delta_{\gamma]}^{\delta]}
-Ph^{[\alpha}_{[\beta}\delta^{\delta]}_{\gamma]}+(\rho-3P)(\frac{1}{3}
\delta^{\alpha}_{[\beta}\delta^{\delta}_{\gamma]}-
\delta^{[\alpha}_{[\beta}\delta^{\delta]}_{\gamma]})\right],\\\nonumber
\mathcal{R}^{\alpha\delta}_{(II)\beta\gamma}&=&\frac{2}{\Phi}\left[
\Pi^{[\alpha}_{[\beta}\delta^{\delta]}_{\gamma]}
+\mathfrak{q}\left(\mathfrak{u}^{[\alpha}\mathfrak{s}_{[\beta}\delta^{\delta]}_{\gamma]}
+\mathfrak{s}^{[\alpha}\mathfrak{u}_{[\beta}\delta^{\delta]}_{\gamma]}\right)\right],\\\nonumber
\mathcal{R}^{\alpha\delta}_{(III)\beta\gamma}&=&4\mathfrak{u}^{[\alpha}\mathfrak{u}_{[\beta}\xi^{\delta]}_{\gamma]}
-\epsilon^{\alpha\delta}_{\lambda}\epsilon_{\beta\gamma\mu}\xi^{\lambda\mu},\\\nonumber
\mathcal{R}^{\alpha\delta}_{(IV)\beta\gamma}&=&\frac{2}{\Phi}\left[\Phi^{[,\alpha}_{[;\beta}
\delta^{\delta]}_{\gamma]}+\frac{\omega_{BD}}{\Phi}\Phi^{[,\alpha}\Phi_{[,\beta}
\delta^{\delta]}_{\gamma]}-\left(\Box\Phi+\frac{\omega_{BD}}{2\Phi}\Phi_{,\lambda}
\Phi^{,\lambda}+\frac{V(\Phi)}{2}\right)\right.\\\nonumber
&\times&\left.\delta^{[\alpha}_{[\beta}\delta^{\delta]}_{\gamma]}\right],
\\\nonumber
\mathcal{R}^{\alpha\delta}_{(V)\beta\gamma}&=&\frac{1}{\Phi}\left[\left(-\frac{\omega_{BD}}
{\Phi}\Phi_{,\lambda}\Phi^{,\lambda}-2V(\Phi)-3\Box\Phi\right)\left(\frac{1}{3}
\delta^{\alpha}_{[\beta}\delta^{\delta}_{\gamma]}-
\delta^{[\alpha}_{[\beta}\delta^{\delta]}_{\gamma]}\right)\right],
\end{eqnarray}
the Riemann tensor is decomposed as
\begin{equation}\label{27}
\mathcal{R}^{\alpha\delta}_{\beta\gamma}=\mathcal{R}^{\alpha\delta}_{(I)\beta\gamma}
+\mathcal{R}^{\alpha\delta}_{(II)\beta\gamma}+\mathcal{R}^{\alpha\delta}_{(III)\beta\gamma}
+\mathcal{R}^{\alpha\delta}_{(IV)\beta\gamma}+\mathcal{R}^{\alpha\delta}_{(V)\beta\gamma}.
\end{equation}
As per Herrera's technique, we introduce the following tensors
\begin{eqnarray*}\label{23}
Y_{\lambda\mu}&=&\mathcal{R}_{\lambda\delta\mu\gamma}\mathfrak{u}^{\delta}
\mathfrak{u}^{\gamma},\\\label{25}
X_{\lambda\mu}=^{*}\mathcal{R}^{*}_{\lambda\delta\mu\gamma}\mathfrak{u}^{\delta}
\mathfrak{u}^{\gamma}&=&
\frac{1}{2}\eta_{\lambda\delta}^{\alpha\epsilon}R^{*}_{\alpha\epsilon\mu\gamma}
\mathfrak{u}^{\delta}\mathfrak{u}^{\gamma},
\end{eqnarray*}
which are computed for the non-static cylinder as
\begin{eqnarray}\nonumber
X_{\lambda\mu}&=&\frac{1}{\Phi}\left(\frac{\rho
h_{\lambda\mu}}{3}+\frac{\Pi_{\lambda\mu}}{2}\right)-\xi_{\lambda\mu}
+\frac{1}{2\Phi}(\Phi_{,\lambda;\delta}h^{\delta}_{\mu}+
\frac{\omega_{BD}}{2\Phi}\Phi_{\lambda}\Phi_{\delta}h^{\delta}_{\mu})\\\label{32}
&+&\frac{h_{\lambda\mu}}{4\Phi}(\Box\Phi+7V(\Phi)),\\\nonumber
Y_{\lambda\mu}&=&\frac{1}{\Phi}\left(\frac{(\rho+3P)h_{\lambda\mu}}{6}+
\frac{\Pi_{\lambda\mu}}{2}\right)+\xi_{\lambda\mu}+\frac{1}{2\Phi}(-\Phi_{,\lambda;\mu}
-\Phi_{,\lambda;\delta}\mathfrak{u}_{\mu}\mathfrak{u}^{\delta}\\\nonumber
&-&\Phi_{,\delta;\mu}\mathfrak{u}_{\lambda}\mathfrak{u}^{\delta}
+\Phi_{,\gamma;\delta}\mathfrak{u}_{\gamma}\mathfrak{u}^{\delta}g_{\lambda\mu})
+\frac{\omega_{BD}}{2\Phi^2}(-\Phi_{,\lambda}\Phi_{,\mu}
-\Phi_{,\lambda}\Phi_{,\delta}\mathfrak{u}^\delta
\mathfrak{u}_\lambda\\\nonumber
&-&\Phi_{,\delta}\Phi_{,\mu}\mathfrak{u}^\delta
\mathfrak{u}_\lambda-\Phi_{,\gamma}\Phi_{,\delta}\mathfrak{u}^{\gamma}\mathfrak{u}^\delta
g_{\lambda\mu})+\frac{h_{\lambda\mu}}{6\Phi}
\left(\frac{\omega_{BD}}{\Phi}\Phi_{,\delta}\Phi^{,\delta}-V(\Phi)
\right).\\\label{33}
\end{eqnarray}
Here,
$\mathcal{R}^{*}_{\lambda\mu\delta\gamma}=\frac{1}{2}\eta_{\alpha\epsilon\delta\gamma}
\mathcal{R}_{\lambda\mu}^{\alpha\epsilon}$ and
$^{*}\mathcal{R}_{\lambda\mu\delta\gamma}=\frac{1}{2}\eta_{\lambda\mu\alpha\epsilon}
\mathcal{R}_{\delta\gamma}^{\alpha\epsilon}$ are the right and left
duals, respectively.

The required scalar quantities are obtained by splitting the tensors
$X_{\lambda\mu}$ and $Y_{\lambda\mu}$ in the following form
\begin{eqnarray*}
X_{\lambda\mu}&=&\frac{X_T}{3}h_{\lambda\mu}+X_{<\lambda\mu>},\\
Y_{\lambda\mu}&=&\frac{Y_T}{3}h_{\lambda\mu}+Y_{<\lambda\mu>},
\end{eqnarray*}
where
\begin{eqnarray*}
X_T&=&X^{\lambda}_{\lambda},\quad
X_{<\lambda\mu>}=h^{\alpha}_{\lambda}
h^{\beta}_{\mu}\left(X_{\alpha\beta}
-\frac{X^{\lambda}_{\lambda}}{3}h_{\alpha\beta}\right),\\
Y_T&=&Y^{\lambda}_{\lambda},\quad
Y_{<\lambda\mu>}=h^{\alpha}_{\lambda}h^{\beta}_{\mu}\left(Y_{\alpha\beta}
-\frac{Y^{\lambda}_{\lambda}}{3}h_{\alpha\beta}\right).
\end{eqnarray*}
The structure scalars incorporating the essential characteristics of
the anisotropic fluid turn out to be
\begin{eqnarray}\nonumber
X_{T}&=&X_{T}^{\text{(m)}}+X_{T}^{\Phi}=\frac{1}{\Phi}(\rho)
+\frac{1}{2\Phi}\left(\frac{5}{2}\Box\Phi
+\Phi_{,\lambda;\alpha}\mathfrak{u}^{\lambda}\mathfrak{u}^{\alpha}
+\frac{\omega_{BD}}{2\Phi}(\Phi_{,\lambda}\Phi^{,\lambda}\right.\\\label{35}
&+&\left.\Phi_{,\lambda}\Phi_{,\alpha}\mathfrak{u}^{\alpha}\mathfrak{u}^{\lambda}
+\frac{21}{2}V(\Phi)\right),\\\nonumber
X_{TF}&=&X_{TF}^{\text{(m)}}+X_{TF}^{\Phi}=-\frac{1}{\Phi}(\frac{\Pi}{2}+\varepsilon\Phi)
+\frac{1}{2\Phi}
\left(\Box\Phi+\Phi_{,\alpha;\lambda}\mathfrak{u}^{\alpha}\mathfrak{u}^{\lambda}\right.\\\label{36}
&+&\left.\frac{\omega_{BD}}
{2\Phi}(\Phi_{,\alpha}\Phi^{,\alpha}+\Phi_{,\lambda}\Phi_{,\alpha}\mathfrak{u}^{\alpha}\mathfrak{u}^{\lambda})\right),\\\nonumber
Y_{T}&=&Y_{T}^{\text{(m)}}+Y_{T}^{\Phi}=\frac{1}{2\Phi}(\rho+3p_r-2\Pi)
-\frac{1}{2\Phi}\left(\Box\Phi
+\Phi_{,\gamma;\alpha}\mathfrak{u}^{\gamma}\mathfrak{u}^{\alpha}\right.\\\label{37}
&+&\left.\frac{\omega_{BD}}{\Phi}(\Phi_{,\gamma}\Phi_{,\alpha}
\mathfrak{u}^{\gamma}\mathfrak{u}^{\alpha})+V(\Phi)\right),\\\nonumber
Y_{TF}&=&Y_{TF}^{\text{(m)}}+Y_{TF}^{\Phi}=\frac{1}{\Phi}(\varepsilon\Phi-\frac{\Pi}{2})
-\frac{1}{2\Phi} \left(\Box\Phi+\frac{\omega_{BD}}
{\Phi}(\Phi_{,\alpha}\Phi^{,\alpha}\right.\\\label{38}
&+&\left.\Phi_{,\gamma}\Phi_{,\beta}\mathfrak{u}^\gamma
\mathfrak{u}^\beta)+\Phi_{,\gamma;\lambda}\mathfrak{u}^{\gamma}\mathfrak{u}^{\lambda}\right).
\end{eqnarray}
The principal stresses and total energy density of the dynamical
system is controlled by the scalars $Y_T$ and $X_T$, respectively.
Moreover, the remaining scalars govern the local anisotropy of the
fluid in the presence of the massive scalar field. Furthermore,
$Y_{TF}$ determines the evolution of the cylinder filled with
anisotropic and inhomogeneous fluid as
\begin{eqnarray}\nonumber
Y_{TF}&=&T_2^{2(\text{eff})}-T_1^{1(\text{eff})}-\frac{1}{2Z^3}\int_0^r
Z'Z^3(D_R T_0^{0(\text{eff})}+\frac{3U}{\Phi
EZ}(\mathfrak{q}-\frac{T_{01}^{\Phi}}{XY}))dr\\\label{39}
&+&\frac{1}{2\Phi}\left[\frac{3\dot{\Phi}}{X^2}\frac{\dot{Z}}{Z}-\frac{2\Phi''}{Y^2}
-\frac{3\Phi'}{Y^2}\frac{Z'}{Z}+\frac{\omega_{BD}}{\Phi}\frac{\Phi'^2}{Y^2}\right].
\end{eqnarray}

\section{Complexity and Evolution of the System}

According to the definition put forward by Herrera \cite{13*}, the
complexity of a configuration depends on its different physical
characteristics. The description of a less complex structure (such
as dust or vacuum) requires fewer matter variables as compared to a
more complex fluid (such as perfect fluid distribution). Thus, in
order to determine the complexity of anisotropic, inhomogeneous and
dissipative self-gravitating structures we require a scalar function
incorporating these essential features. Equation (\ref{39}) shows
that the structure scalar $Y_{TF}$ includes the effects of
complexity inducing features as well as the massive scalar field.
Moreover, $Y_{TF}$ has been treated as a suitable candidate of
complexity factor in the case of a static regime \cite{100a}. Thus,
$Y_{TF}$ can adequately represent the complexity of the dynamical
cylinder. Furthermore, the pattern of evolution is an important
aspect of non-static celestial systems. In this section, we discuss
two possible evolution patterns of the anisotropic fluid.

\subsection{The Homologous Evolution}

If the density of the collapsing configuration is the same
throughout, i.e., matter falls into the core at the same rate in the
internal region then the celestial body evolves homologously.
However, if the velocity with which the matter collapses is not
proportional to the radial distance then the setup follows a
non-homologous pattern. Employing Eqs.(\ref{7}) and (\ref{57}), the
heat flux is expressed as
\begin{equation}\label{58}
\frac{1}{2E\Phi}\left(\mathfrak{q}-\frac{T_{01}^\Phi}{XY}\right)
=\frac{1}{3}D_R(\Theta-\sigma) -\frac{\sigma}{Z},
\end{equation}
which leads to
\begin{equation}\label{62}
D_R\left(\frac{U}{Z}\right)=\frac{1}{2E\Phi}
\left(\mathfrak{q}-\frac{T_{01}^{\Phi}}{XY}\right)+\frac{\sigma}{Z}.
\end{equation}
Consequently, the velocity of the collapsing cylinder is obtained as
\begin{equation}\label{63}
U=Z\int_0^rZ'\left[\frac{1}{2E\Phi}
\left(\mathfrak{q}-\frac{T_{01}^{\Phi}}{XY}\right)+\frac{\sigma}{Z}\right]dr+f(t)Z,
\end{equation}
where $f(t)=\frac{U_{\Sigma}}{Z_{\Sigma}}$ is an integration
function. The condition of homologous evolution ($U\sim Z$
\cite{110a}) is obtained if the integral in the above equation
vanishes. Thus, the condition
\begin{equation}\label{64}
\frac{1}{2E\Phi}
\left(\mathfrak{q}-\frac{T_{01}^{\Phi}}{XY}\right)+\frac{\sigma}{Z}=0,
\end{equation}
must hold for homologous evolution. This condition suggests that for
two shells of fluids, the ratio of the aerial radii is constant. We
proceed by considering $Z(t,r)$ as a separable function of $t$ and
$r$.

\subsection{The Homogeneous Expansion}

Another simplest pattern of evolution is the homogeneous expansion
which corresponds to $\Theta'=0$. In other words, if the rate at
which the self-gravitating system evolves is independent of the
radial co-ordinate then the system collapses or expands
homogeneously. Applying the homogeneous condition to Eq.(\ref{58})
yields
\begin{equation}\nonumber
\frac{1}{2E\Phi}\left(\mathfrak{q}-\frac{T_{01}^\Phi}{XY}\right)
=-\frac{1}{3}D_R(\sigma) -\frac{\sigma}{Z}.
\end{equation}
If the fluid is homologous as well then the homogeneous condition
reduces to $D_R \sigma=0$ or $\sigma=0$ (because of the regularity
conditions at $r=0$). Thus, Eq.(\ref{58}) can be written as
\begin{equation}\label{65}
\mathfrak{q}=\frac{T_{01}^\Phi}{XY}.
\end{equation}
It must be noted that under the influence of the scalar field,
homogeneous expansion does not lead to a dissipation-free matter
source. However, in GR, a matter distribution with
$\sigma=\Theta'=0$ is non-dissipative as well as homologous.

\section{Kinematical and Dynamical Variables}

In this section, we choose the simplest pattern of evolution based
on the analysis of some important kinematical quantities. If a fluid
follows the homologous pattern of evolution then Eq.(\ref{58})
provides
\begin{equation}\nonumber
(\Theta-\sigma)'=\left(\frac{3\dot{Z}}{XZ}\right)'=0.
\end{equation}
Imposing the condition $Z(t,r)=Z_1(t)Z_2(r)$ implies that $X'=0$.
Consequently, the fluid is geodesic ($a=0$). Without loss of
generality, we consider $X=1$. Conversely, consider the geodesic
condition
\begin{equation*}
\Theta-\sigma=\frac{3\dot{Z}}{Z}.
\end{equation*}
Successively differentiating the above equation with respect to $r$
and assuming that $(\Theta-\sigma)'$ equals its Taylor series close
to the center leads to a homologous fluid \cite{13*}. Thus, the
cylindrical object evolves homologously if and only if the fluid is
geodesic. For this reason, we consider the homologous pattern as the
simplest pattern of evolution. If the self-gravitating structure is
non-dissipative then the shear tensor under the effect of the
massive scalar field takes the form
\begin{equation*}
\sigma=\frac{ZT_{01}^{\Phi}}{2Z'}.
\end{equation*}

If the celestial system adopts the mode of homogeneous expansion,
then the non-dissipative scenario produces $T_{01}^{\Phi}=0$.
Further, the shear scalar is computed as
\begin{eqnarray*}
\sigma&=&\frac{1}{2Z^3}\int_0^r\frac{Z^3}{X}T_{01}^{\Phi}dr+\frac{h(t)}{Z^3}=\frac{h(t)}{Z^3},
\end{eqnarray*}
where $h(t)$ is an arbitrary integration function. Since
$Z\rightarrow0$ as $r\rightarrow0$, therefore, $h(t)$ must approach
to zero. It is deduced that homogeneous expansion implies homologous
evolution when $\mathfrak{q}=0$ (since $\sigma=0\Rightarrow U\sim
Z$). On the other hand, if the non-dissipative fluid evolves
homologously then $\sigma=\frac{ZT_{01}^{\Phi}}{2Z'}$ which implies
\begin{equation*}
\Theta'=(\frac{ZT_{01}^{\Phi}}{2Z'})'.
\end{equation*}
Thus, homologous evolution does not imply homogeneous expansion. The
C-energy of a homologously evolving cylinder is related to the rate
of collapse as
\begin{equation}\label{105}
D_TU=-\frac{m}{Z^2}-\frac{Z}{2}T_1^{1(\text{eff})}+\frac{1}{8Z},
\end{equation}
which leads to
\begin{eqnarray}\nonumber
\frac{3D_TU}{Z}&=&\frac{1}{2}(T_0^{0(\text{eff})}-T_1^{1(\text{eff})}+2T_2^{2(\text{eff})})
+Y_{TF}-\frac{1}{2\Phi}\left[3\dot{\Phi}\frac{\dot{Z}}{Z}-\frac{2\Phi''}{Y^2}
-\frac{3\Phi'}{Y^2}\frac{Z'}{Z}\right.\\\label{106}
&+&\left.\frac{\omega_{BD}}{\Phi}\frac{\Phi'^2}{Y^2}\right].
\end{eqnarray}
Employing the field equations and the definition of $U$ in the above
relation, the complexity factor is expressed as
\begin{equation}\nonumber
Y_{TF}=\frac{\ddot{Z}}{Z}-\frac{\ddot{Y}}{Y}+\frac{1}{2\Phi}\left[3\dot{\Phi}\frac{\dot{Z}}{Z}-\frac{2\Phi''}{Y^2}
-\frac{3\Phi'}{Y^2}\frac{Z'}{Z}+\frac{\omega_{BD}}{\Phi}\frac{\Phi'^2}{Y^2}\right].
\end{equation}
In the next sections, we utilize the constraints corresponding to
vanishing complexity and homologous fluid to obtain solutions for
$\mathfrak{q}=0$ and $\mathfrak{q}\neq0$ by assuming  $X=1$. Since
the number of unknowns exceeds the number of equations, we consider
the massive scalar field as
\begin{equation}\label{107}
\Phi(t,r)=\Phi(t)=\Phi_0t^m,
\end{equation}
where $\Phi_0$ is the present day value of the scalar field and $m$
is a constant. Moreover, for this choice of the scalar field
$T_{01}^{\Phi}=0$. Thus, the homogeneous and homologous evolution
conditions are same for $\mathfrak{q}=0$ which implies a unique mode
of evolution.

\subsection{Case 1: Non-dissipative Fluid}

In the non-dissipative scenario, the homologous condition becomes
\begin{equation}\label{66}
Y(t,r)=Z(t,r)h_1(r),
\end{equation}
where $h_1(r)$ is an arbitrary function of integration. The wave
equation and condition of vanishing complexity corresponding to
Eqs.(\ref{107}) and (\ref{66}) generate the following relations
\begin{eqnarray*}
&&\frac{\Phi_0t^{m-1}}{(2
\omega_{BD}+3)h_1(r)ZZ'}\left(-4tZ\left(h_1(r)^3\left(Z'
\left(m(\omega_{BD}+1)\dot{Z}+t\ddot{Z}\right)+t\dot{Z}\dot{Z}'\right)\right.\right.\\
&&\left.\left.+th_1'(r)\right)+h_1(r)^3 Z^2 \left(Z'\left((m-2)
m\omega_{BD}+t^2V'(\Phi)\right)-2t\left(m(\omega_{BD}+1)
\dot{Z}'\right.\right.\right.\\
&&\left.\left.\left.+t\ddot{Z}'\right)\right)-2t^2h_1(r)Z'\left(h_1(r)^2\dot{Z}^2-1\right)\right)=0,\\
&&V(\Phi)=\frac{\Phi_0 t^{m-2}}{h_1(r)^3
ZZ'}\left(2t^2h_1(r)^3\dot{Z}
\dot{Z}'+4t^2h_1(r)^3Z'\ddot{Z}+m^2\omega_{BD}h_1(r)^3ZZ'\right.\\
&&\left.+2m^2h_1(r)^3ZZ'-2m h_1(r)^3ZZ'-m th_1(r)^3Z'\dot{Z}+2t^2
h_1'(r)\right).
\end{eqnarray*}
A suitable choice of $h_1(r)$ determines the complete solution.

\subsection{Case 2: Dissipative Fluid}

In the non-dissipative case, the homologous, wave equation and
complexity-free condition, respectively, read
\begin{eqnarray*}
&&Y=h_2(r)\exp\left(\int_1^t\frac{\Phi_0 t^{m}Z\dot{Z}'-Z'
\dot{Z}}{\left(\Phi_0t^{m}-1\right)ZZ'}\,dt\right),\\
&&\frac{\Phi_0t^{m-1}}{(2\omega_{BD}+3)YZ}\left(-4t^2Y'ZZ'-2tY^2Z\left(2t\dot{Y}
\dot{Z}+Z\left(m(\omega_{BD}+1)\dot{Y}+t
\ddot{Y}\right)\right)\right.\\
&&\left.+Y^3\left(Z^2\left((m-2)m
\omega_{BD}+t^2V'(\Phi)-2(t\dot{Z})^2-4tZ\left(m
(\omega_{BD}+1)\dot{Z}\right.\right.\right.\right.\\
&&\left.\left.\left.\left.+t\ddot{Z}\right)\right)\right)+2t^2Y\left(Z'^2+2Z
Z''\right)\right)=0,\\
&&V(\Phi)=\frac{\Phi_0t^{m-2}}{Y^3
Z}\left(2t^2\dot{Y}Y^2\dot{Z}+2t^2Y' Z'-2t^2YZ''+Y^3
\left(t\left(4t\ddot{Z}-m\dot{Z}\right)\right.\right.\\
&&\left.\left.+m (m(\omega_{BD}+2)-2)Z\right)\right),
\end{eqnarray*}
where $h_2(r)$ is an integration function whose appropriate form
completely specifies the above system of equations for the chosen
scalar field.

\section{Stability of $Y_{TF}=0$ Condition}

It is possible that a system that initiates with complexity-free
interior develops complex nature at a later time, i.e., the
condition of vanishing complexity may be disturbed during the
evolution of the system. In this section, we investigate if the
vanishing complexity condition can sustain throughout the homologous
evolution of matter configuration corresponding to
$\Phi(t,r)=\Phi(t)=\Phi_0t^m$. Equations (\ref{100}) and (\ref{15})
determine the evolution of the complexity as
\begin{eqnarray}\label{102}
\dot{Y}_{TF}+\frac{\dot{\Pi}}{\Phi}+\frac{3\dot{Z}}{Z}Y_{TF}+(\rho+P_r)\frac{\sigma}{2\Phi}
+\frac{1}{2Y\Phi}(\mathfrak{q}'-\frac{\mathfrak{q}Z'}{Z})+\frac{2\Pi
\dot{Z}}{Z\Phi}+S_1=0,
\end{eqnarray}
where the term $S_1$ contains the effects of scalar field and is
given as
\begin{eqnarray*}
S_1&=&\frac{(T_1^{1\Phi}-T_2^{2\Phi})^{.}}{2\Phi}-\frac{(T_0^{1\Phi})'}{\Phi}
-\frac{(T_0^{1\Phi})'}{2\Phi}(\frac{Y'}{Y}-\frac{Z'}{Z})-
\frac{(T_0^{0\Phi}-T_1^{1\Phi})\dot{Y}}{2Y\Phi}\\
&-&\frac{5(T_0^{0\Phi}-T_2^{2\Phi})\dot{Z}}{2Z\Phi}-\dot{Y}_{TF}^{\Phi}-Y_{TF}.
\end{eqnarray*}

We first consider $\mathfrak{q}=0$ with $\Pi=\sigma=Y_{TF}=0$ at
$t=0$. Equation (\ref{102}) and its derivative with respect to $t$
are, respectively expressed as
\begin{eqnarray}\label{103}
S_1&=&-(\dot{Y}_{TF}+\dot{\Pi}),\\\label{104}
\ddot{Y}_{TF}+\frac{\ddot{\Pi}}{\Phi}-\frac{\dot{\Pi}\dot{\Phi}}{\Phi^2}&=&3S_1
-\dot{S_1}+\frac{\dot{\Pi}\dot{Z}}{Z\Phi},
\end{eqnarray}
which leads to the following forms of first and second
$t$-derivatives of Eq.(\ref{39})
\begin{eqnarray*}
S_1+3\left(\frac{\dot{\Phi}\dot{Z}}{Z}\right)^{.}&=&\frac{\partial}{\partial
t}\left(\int_0^rZ^3(T_0^{0\text{(eff)}})'dr\right),\\
3S_1-\dot{S}_1+\frac{\dot{\Pi}\dot{Z}}{Z\Phi}-3\left(\frac{\dot{\Phi}\dot{Z}}{Z}\right)^{..}
&=&\frac{\partial^2}{\partial
t^2}\left(\int_0^r-Z^3(T_0^{0\text{(eff)}})'dr\right).
\end{eqnarray*}
The higher order derivatives can be obtained by proceeding in the
same manner. It is observed that the state variables (anisotropic
pressure and inhomogeneous energy density) along with the scalar
field induce complexity in the system. Thus, the stability of
$Y_{TF}=0$ condition depends on these factors. For
$\mathfrak{q}\neq0$, heat dissipation also affects the condition of
zero complexity.

\section{Summary}

Self-gravitating systems are complicated and intriguing cosmic
objects. Researchers examine the origin and evolution of these
systems to gain insight into the structure of the universe. In this
paper, we have formulated a complexity factor to investigate the
relations between different state parameters of a dynamical cylinder
in the context of MBD gravity. The complexity of the dissipative
setup has been determined through structure scalars obtained via the
orthogonal splitting of the Riemann tensor. The dynamics of the
cylindrical regime have been incorporated in the definition of
complexity by considering the pattern of evolution. We have examined
two evolution modes namely, homologous and homogeneous. Solutions
corresponding to $\mathfrak{q}=0$ and $\mathfrak{q}\neq0$ have been
developed by applying the homologous and zero complexity conditions.
Finally, the criteria under which the self-gravitating cylinder
departs from the initial state of zero complexity have also been
discussed.

The splitting of the Riemann tensor has yielded structure scalars
that govern the mechanism and inhomogeneous structure of the
anisotropic system. It has been noted that heat dissipation,
anisotropy and inhomogeneity are the main factors contributing to
the complexity of the non-static model in GR. However, the presence
of massive scalar field and potential function in structure scalars
indicates that complexity of the MBD model depends on the scalar
field as well. Thus, the cylindrical system in GR is less
complicated as compared to its MBD analog. We have chosen the
structure scalar $Y_{TF}$ to represent the complexity of the system
based on the following reasons.
\begin{itemize}
\item It incorporates the effects of anisotropic pressure, heat
dissipation and inhomogeneous energy density of the configuration.
\item The complexity of the static cylinder has been adequately
measured through $Y_{TF}$ \cite{100a}. Thus, choosing $Y_{TF}$ for
the dynamical system ensures that the current definition can be
restored in the static regime.
\end{itemize}
Furthermore, the homologous pattern of evolution fulfils the
condition of geodesic fluid and conversely the geodesic fluid
evolves homologously. Therefore, we have chosen the homologous
pattern as the least complex mode of evolution. It is noted that the
scalar field influences the evolution of the system. Consequently,
the condition $\mathfrak{q}=0$ does not lead to a shear-free
self-gravitating model. The implementation of vanishing complexity
and homologous pattern for $\Phi(t,r)=\Phi(t)=\Phi_0t^m$ leads to
open systems corresponding to non-dissipative as well as dissipative
scenarios. Suitable choices of integration functions specify the
system completely. Moreover, a perturbation in the scalar field or
state determinants (heat flux, pressure, density) may disturb the
system from its state of zero complexity. It is noteworthy that if
$\Phi=\text{constant}$ and $\omega_{BD}\rightarrow \infty$ then all
the results derived in this paper reduce to their GR counterparts.

\end{document}